\documentclass[preprint,showpacs,preprintnumbers,amsmath,amssymb]{revtex4}

\usepackage{graphicx,subfigure}
\usepackage{dcolumn}
\usepackage{bm}
\usepackage{psfrag}

\newcommand{\be}{\begin{equation}}
\newcommand{\ee}{\end{equation}}
\newcommand{\bea}{\begin{eqnarray}}
\newcommand{\eea}{\end{eqnarray}}

\begin{document}

\bibliographystyle{apsrev}

\preprint{UAB-FT-568}

\title{On the Gravitational Field of Antimatter} 

\author{Eduard Mass{\'o}}
\author{Francesc Rota}

\affiliation{Grup de F{\'\i}sica Te{\`o}rica and Institut
de F{\'\i}sica d'Altes
Energies\\Universitat Aut{\`o}noma de Barcelona\\
08193 Bellaterra, Barcelona, Spain}


\date{\today}

\begin{abstract}
The gravitational field of matter and that of antimatter could differ. This might be one signature of quantum gravity. We show that primordial Big Bang Nucleosynthesis restricts such a possibility.
\end{abstract}

\maketitle


The symmetry between matter and antimatter arises naturally in 
conventional quantum
field theory. With the priors of Lorentz invariance, hermiticity, and locality,
the theory is CPT conserving. Among other consequences, CPT conservation
implies that the inertial masses of a particle and its corresponding antiparticle are
equal. When technically feasible, this identity among masses
has been checked in the laboratory. General relativity is based on
the equality of inertial and gravitational masses, and this
hypothesis should apply in principle both to matter and to
antimatter. However, it is not at all obvious that this assumption 
will be tenable when going from classical general relativity to
a quantum theory of gravitation. The path to quantum gravity is not yet 
established,
but one thing we know for sure is that we have to go beyond conventional 
field theory;
it follows then that we cannot invoke the kind of matter-antimatter symmetry
we alluded to. In fact, there is the claim 
\cite{scherk,goldman,ellis,ahluwalia}
about the possibility of a fundamental asymmetry between matter and
antimatter that could manifest as a signature of quantum gravity
(for a review see \cite{Nieto:xq} and references therein).

It is not a trivial issue to probe the gravitational properties of
antimatter and compare them with those of matter. Apart from our own
work, that we will present below, we are aware of only one instance
where this comparison has been possibly done \cite{losecco,pakvasa}. 
In the supernova
1987A collapse, a huge flux of neutrinos and antineutrinos was emitted.
The flux, when crossing the Earth, left a signal in underground detectors.
The statistical analysis concluded that \cite{pakvasa2}, 
at the 90\% confidence level or more, 
at least one $\nu_e e$ event was detected, the rest being $\bar \nu_e p$.
Assuming $\nu_e$ and $\bar \nu_e$ were indeed detected, 
and taking into account that all events were recorded in about a 10 second
period, one can put a restriction on the geodesic deviation of
these particles when following a path from the 
supernova in the Large Magellanic Cloud (LMC) until the Earth.
This is expressed conveniently in terms of the 
post-Newtonian parameter $\gamma$ ($\gamma=1$ 
for all particles and antiparticles in Einstein's theory).
One gets \cite{losecco,pakvasa}
\be
|\gamma_{\nu_e} \, - \, \gamma_{\bar \nu_e}  | < 10^{-5} - 10^{-6}
\label{nu}
\ee
The bound (\ref{nu}) 
on the (weak) equivalence principle for matter and antimatter 
is subject to some caveats. As we said, it is necessary that at least one
of the detected events corresponds to a $\nu_e$. In addition, if the violation
of the equivalence principle is due to new long-range forces, then the limit
(\ref{nu}) is relaxed if the range is shorter than the LMC-Solar System
distance. Also, screening effects might invalidate (\ref{nu}), 
as is discussed in \cite{raffelt}.

We should also say that, after the successful antihydrogen production 
\cite{Baur:1995ck,Amoretti:2002um}, 
a laboratory limit on conceivable differences between the fall of 
hydrogen and antihydrogen in the Earth gravitational field will be 
in principle possible \cite{web}.

This last experimental proposal and the limit (\ref{nu}) are concerned
with the effects of a gravitational field on matter and antimatter.
We would also like 
to compare the gravitational field produced by
bulk matter with the corresponding gravitational field produced by
antimatter. This is very hard of course because we need a large macroscopic
sample of antimatter and measure the gravitational field it produces.
Fortunately,
the very early universe offers a ``laboratory'' for such a test. 
To start with, one has 
comparable densities of matter and
antimatter. Also, the particle and antiparticle energy densities are responsible for the gravitational Hubble expansion.
In the standard Big Bang model, the expansion rate of the universe, measured
by the Hubble parameter $H$, is given by the Friedmann equation
in terms of the energy densities $\rho_i$,
\be
H^2 = \frac{8\pi}{3}\, G_N\,  \sum_i \rho_i
\label{friedmann}
\ee
where $G_N$ is the Newton's gravitational constant. In (\ref{friedmann}),
the contribution of curvature and cosmological constant has been neglected. 
This is justified, since we shall
apply the equation in the very early universe, specifically in the period of
Big Bang Nucleosynthesis (BBN).

We see that in (\ref{friedmann}) the densities $\rho_i$
of the different species $i$ present in the early universe
contribute to the expansion with the same strength
$G_N$, so that, as a consequence, matter and antimatter enter with the same
gravitational constant.
The root for that is found in Einstein equations
\be
R^{\mu\nu}-\frac{1}{2}\, g^{\mu\nu} R =  8\pi\, G_N T^{\mu\nu}
\label{einstein}
\ee
The Friedmann equation (\ref{friedmann}) can be deduced from 
Einstein equations, if we
use the Robertson-Walker metric to calculate
the Ricci tensor $R^{\mu\nu}$ and the Ricci scalar $R$,
and describe
the energy-momentum tensor $T^{\mu\nu}$ as a perfect fluid. 
The energy densities $\rho_i$ are
contained as a perfect fluid source in $T^{\mu\nu}$ and they contribute
with the same strength $G_N$.

If matter and antimatter produce different gravitational fields,
we would expect a violation of Einstein equations (\ref{einstein}),
and as a consequence the Friedmann equation (\ref{friedmann})
 would also be invalid.
Even if we do not have a fundamental theory in this case,
we shall proceed in the following way to constrain such
matter/antimatter asymmetry.

We will make use of the fact that the Friedmann equation
gives a fair description \cite{carroll}
of the universe expansion in the BBN period.
It is reasonable to think that if the Friedmann equation (\ref{friedmann})
is to be altered, the modifications are bound to be not excessively drastic. 

With this argument in mind,
we shall postulate a minimally modified Friedmann equation where matter and
antimatter enter with different strengths,
\be
H^2 = \frac{8\pi}{3}\, [ G_N\, \rho_M \, + \,  G_N (1+\delta_\gamma)\,
\rho_\gamma  \, + \, G_N (1+\delta_A)\, \rho_A ]
\label{friedmann2}
\ee
Several comments are in order. We have left $G_N$ multiplying the energy 
density matter
contribution $\rho_M$, since $G_N$ is the usual magnitude measured for
the gravitational field produced by a source of matter.
The energy density of electrons and neutrinos will  
contribute to $\rho_M$ (protons and neutrons are present in the BBN era 
but, in practise, it is the relativistic particles that dominate $\rho_M$).
A modification appears for antimatter, which enters our equation
(\ref{friedmann2}) with a strength $G_A=G_N (1+\delta_A)$. 
We will have positrons
and antineutrinos contributing to the antimatter energy density $\rho_A$.
Also, there are photons with an energy density $\rho_\gamma$
that contribute to the universe expansion. 
Once we admit the possibility of breaking the matter/antimatter symmetry,
we should let the photons enter the modified Friedmann equation with
yet another gravitational constant $G_\gamma=G_N (1+\delta_\gamma)$
different in principle from both matter and antimatter.

We shall apply BBN arguments to constrain $\delta_A$ and $\delta_\gamma$. BBN has already quite a long list of limits to different models \cite{Sarkar}. Let us first summarise how it works in our case.

In the standard BBN, it is crucial when weak interactions
freeze out, since from this moment on the interconversion between
protons and neutrons is ineffective, and primordial nucleosynthesis
starts. A modification of the expansion rate $H$ makes the freeze out
temperature different from the standard one and this affects the prediction
of the primordial yields of light elements. We will calculate the
theoretical yields using the modified Friedmann equation (\ref{friedmann2})
with the parameters $\delta_A$ and $\delta_\gamma$. Then we will compare the theoretical predictions
with observational data and require agreement. In this
way we shall obtain bounds on the parameters $\delta_A$ and $\delta_\gamma$. 
For the
predictions we need to specify the number of neutrinos $N_\nu$ and 
the neutrino/antineutrino asymmetry which is described by
the chemical potential $\mu_\nu$. 
This is easy to understand: changing the number of neutrinos
changes the expansion of the universe, on the one hand, and, on the other hand, 
$\mu_\nu$ enters in the proton/neutron interconversion reaction
rates. We shall assume 
the values of the standard BBN:  $N_\nu=3$, as collider measurements
tell us, and  $\mu_\nu =0$. In the standard BBN analysis one gets
that $\mu_\nu/T$ has to be small \cite{steigman}, with the electron
neutrino asymmetry more restricted that the others. Neutrino oscillations
seem to imply that the limits for the electron neutrino apply also
to the other neutrinos \cite{dolgov}. Even if we work in non-standard
BBN, we think it is reasonable to set $\mu_\nu =0$ in our analysis, and in
this way we do not have an extra parameter to play with. 

In addition, the predictions depend on the baryon number density. So, still another parameter $\eta$ which is this density normalised to the photon number density is needed in our calculation. Until very recently BBN was the only way we had to get information on $\eta$. This situation has changed with the observational results \cite{WMAP} on the cosmic microwave background, leading to the allowed range
\begin{equation}
  \label{eta}
  \eta = (6.13 \pm 0.25)\times 10^{-10}
\end{equation}
We will use (\ref{eta}) as an input. Finally, the primordial yields of light elements also depend on the neutron lifetime; we take $\tau_n = 885.7 \pm 0.8$ sec \cite{PDG}. 

The success of BBN in constraining models is of course not only due to the theoretical framework we have but also to the fact that we possess observational data. The data is not without debate since some hypotheses have to be done to go from the observed abundances to the primordial abundances. We adopt the recent measurements coming from \cite{kirkman}, where the deuterium detection in quasar absorption systems at high redshift leads to the following ratio of deuterium to hydrogen by number
\begin{equation}
  \label{deuteri}
  {\rm D/H} = 2.78\,^{+0.44}_{-0.38}\times 10^{-5}
\end{equation}

Unfortunately, the determinations of primordial $^4$He suffer from discrepancies between the different estimates. We prefer to take a conservative attitude and, instead of choosing a specific measurement among all, we adopt a range which encompasses the existing measurements. The observational primordial $^4$He mass fraction is taken to be in the range
\begin{equation}
  \label{heli}
 0.228 \leq Y_P \leq 0.248 
\end{equation}
However we will comment below what would happen if we follow a more recent analysis.

\begin{figure}[htb]
\begin{center}
\includegraphics[width=10cm, height=10cm]{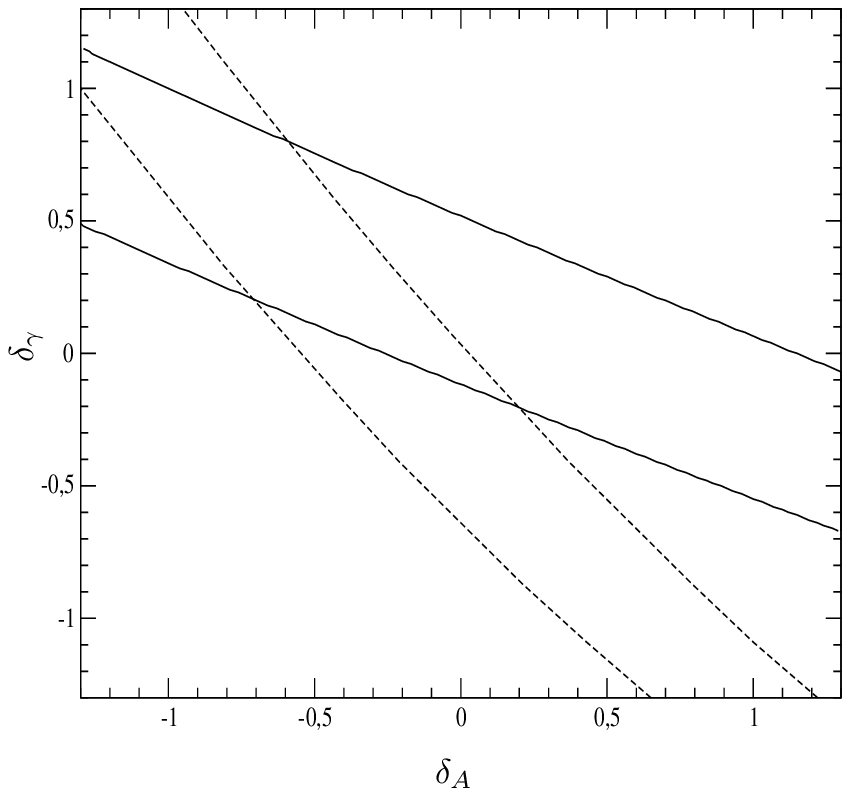}
\end{center}
\caption{\label{fig1} Constraints on the ($\delta_A,\delta_\gamma$) plane coming from the requirement that the predicted abundances are consistent with the observed values. We show the constraints from deuterium as a solid line, and from $^4$He as a dashed line. We have put $\eta = 6.13\times 10^{-10}$.} 
\end{figure}

We are now ready to explain how we actually get our results. We run a modified version of the numerical Kawano code \cite{kawano}. We first introduce as inputs the central values of $\eta$ and $\tau_n$ that we mentioned before. For each value of $\delta_A$ and $\delta_\gamma$ we get a prediction for the abundances of deuterium and helium-4. Comparing the predictions with the data in (\ref{deuteri}) and (\ref{heli}) we get two allowed bands. These are shown in Fig. \ref{fig1}.  Requiring consistency for deuterium leads to the band between the two solid lines, while the dashed lines enclose the allowed band for $^4$He. It is no surprise to get a band since there is some cancellation between the effects of $\delta_A$ and $\delta_\gamma$ if the sign of these two parameters turns out to be opposite. The bands have not the same slope and thus we get an allowed region when we consider both elements, leading to the bounds $-1.4\leq \delta_A \leq 0.2$, $-0.2\leq \delta_\gamma \leq 1.2$. 

\begin{figure}[htb]
  \centering
  \subfigure[]{
   \includegraphics[width=6.5cm, height=6.5cm]{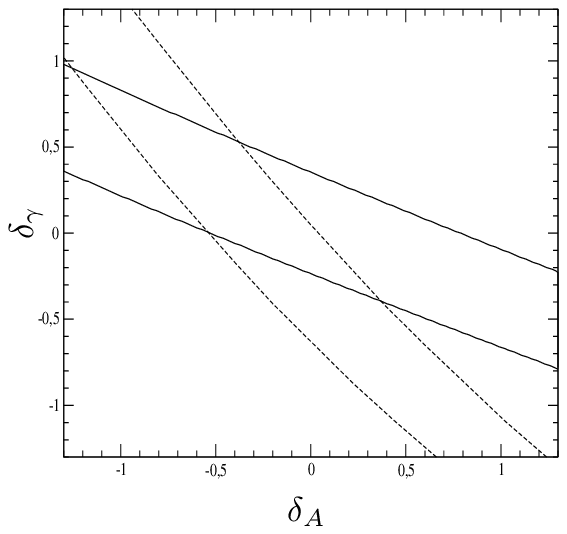}
    \label{fig2}}
  \subfigure[]{
    \includegraphics[width=6.5cm, height=6.5cm]{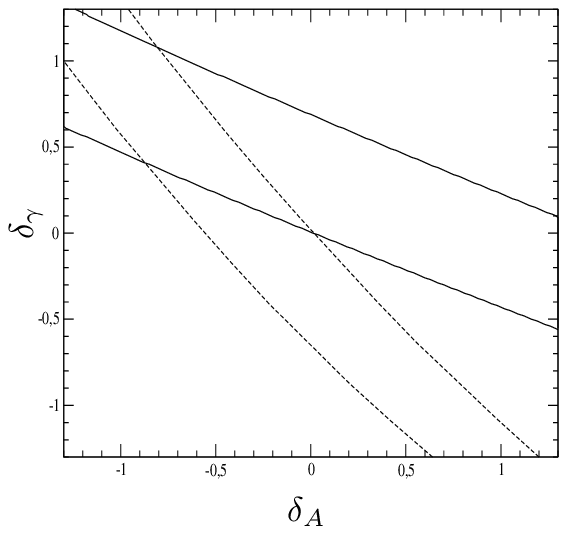}
    \label{fig3}}
  \caption{ Region in the ($\delta_A,\delta_\gamma$) plane where the modified Friedmann equation leads to abundances in agreement with observation. We display the limits coming from deuterium (solid line), and $^4$He (dashed line). The values for the baryon number density are (a) $\eta = 5.88\times 10^{-10}$, and (b) $\eta = 6.38\times 10^{-10}$. }
\end{figure}

We have arrived at our main objective: to show that BBN provides a limit on the possibility that the gravitational field of antimatter and radiation differs from the gravitational field of matter. We got limitations on $\delta_A$ and $\delta_\gamma$ for $\eta = 6.13 \times 10^{-10}$. We should now allow $\eta$ to vary in the experimental error range and strictly speaking we should do that introducing $\eta$ as a statistical variable. We will not do it this way because, as we said, the observational $^4$He, which is one of the main inputs, is taken with a large error to take into account measurements that are not consistent at one $\sigma$ among themselves. We hope the situation will improve in the future, but for the time being we do not undertake the task of treating $\eta$ statistically. For our purposes, we limit ourselves to find the allowed region in the $\delta_A$, $\delta_\gamma$ plane for $\eta = 5.88\times 10^{-10}$ and $\eta = 6.38\times 10^{-10}$, namely the one-$\sigma$ observational extremes coming from WMAP. We keep $\tau_n$ fixed since this does not introduce any appreciable change. The results are shown in Figs. \ref{fig2} and \ref{fig3}. We conservatively take as the allowed range for our parameters the union of the ranges in the figures, which is
\begin{eqnarray}
  \label{delta_range}
-1.6 \leq &\delta_A& \leq 0.4 \nonumber\\
-0.4 \leq &\delta_\gamma& \leq 1.5
\end{eqnarray}
This is the main numerical result of our work.

In our analysis, we have used a $^4$He abundance that is large enough to comprise different measurements. In fact, there are recent analysis leading to values $Y_P$ that are not consistent with both the measured D/H (\ref{deuteri}) and $\eta$ (\ref{eta}). For example, taking $Y_P = 0.238 \pm 0.005$ \cite{barger} leads to this situation. Allowing a non-zero $\delta_A$ and $\delta_\gamma$ would make a consistent scenario where $Y_P$, D/H and $\eta$ agree. We think however that the tension between the observables is likely to be solved by an improvement of the observations. So we regard our work as giving a constraint on antimatter gravity, rather than giving a solution to a crisis.

We finally would like to stress that if in our modification of the Friedmann equation (\ref{friedmann2}) we have some relation between $\delta_A$ and $\delta_\gamma$ then the bounds will be tighter. For example, assume the following modified Friedmann equation
\begin{equation}
  \label{friedmann3}
  H^2 = \frac{8\pi}{3}\, [ G(1+\delta)\, \rho_M \, + \,  G \,
\rho_\gamma  \, + \, G (1-\delta)\, \rho_A ]
\end{equation}
Here matter and antimatter couple differently but the departure from the radiation coupling is somehow symmetric. This is what would be expected in theories with a ``graviphoton'' \cite{zachos} and has been shown to be consistent \cite{goldman2} with energy-conservation arguments \cite{morrison}. In (\ref{friedmann3}) we have to identify $G_N = G(1+\delta)$. The limits from BBN on $\delta$ are
\begin{equation}
-0.02 \leq \delta \leq 0.12
\end{equation}
The bound is now tighter because now we have (\ref{friedmann3}) that has one free parameter instead of (\ref{friedmann2}) with two parameters. We would have another one-parameter model if we put $\delta_\gamma=0$ and let only vary $\delta_A$. The bound would then be 
\begin{equation}
- 0.52 \leq \delta_A \leq 0.05 \hspace{1.5cm} (\delta_\gamma = 0)
\end{equation}
The reverse situation is that $\delta_A = 0$, $\delta_\gamma \neq 0$. Physically, it amounts to say that antimatter and matter behave identically while it is the gravitational field of radiation that is different. Notice that an alternative physical interpretation is that fermions and bosons produce a different gravitational field. In fact, this possibility has been considered in \cite{barrow}, where the authors use BBN to constrain this fermion/boson asymmetric model.

\begin{acknowledgments}
We acknowledge support by the CICYT Research Project FPA2002-00648,
by the EU network on Supersymmetry and the Early Universe
(HPRN-CT-2000-00152), and by the \textit{Departament d'Universitats,
Recerca i Societat de la Informaci{\'o}}, Project 2001SGR00188.
\end{acknowledgments}




\newpage

\end{document}